\documentstyle[12pt]{article}
\def\lsim{\mathrel{\rlap{\lower4pt\hbox{\hskip1pt$\sim$}}
    \raise1pt\hbox{$<$}}}         
\def\gsim{\mathrel{\rlap{\lower4pt\hbox{\hskip1pt$\sim$}}
    \raise1pt\hbox{$>$}}}         
\def\ut#1{$\underline{\smash{\vphantom{y}\hbox{#1}}}$}
\def\overleftrightarrow#1{\vbox{\ialign{##\crcr
    $\leftrightarrow$\crcr
    \noalign{\kern 1pt\nointerlineskip}
    $\hfil\displaystyle{#1}\hfil$\crcr}}}

\begin{document}
\begin{center}

{\bf The Weak Parity-Violating Pion-Nucleon Coupling}
\vspace{.5in}

E.M. Henley \\

{\em Physics Department, FM-15 and Institute for Nuclear Theory, HN-12 \\
University of Washington, Seattle, Washington 98195}

\vspace{2mm}
W-Y.P. Hwang \\

{\em Department of Physics, National Taiwan University \\
Taipei, Taiwan 10764}

\vspace{2mm}
L.S. Kisslinger \\
{\em Department of Physics, Carnegie-Mellon University \\
Pittsburgh, Pennsylvania 15213}

\vspace{.5in}
{\bf Abstract}
\end{center}

We use QCD sum rules to obtain the weak parity-violating pion-nucleon coupling
constant $f_{\pi NN}$.
We find that $f_{\pi NN}\approx 3\times 10^{-8}$,
up to an order of magnitude
smaller than the ``best estimates''
based on quark models.
This result follows from the cancellation between perturbative and
nonperturbative QCD processes not found in quark models, but explicit
in the QCD sum rule method.  Our result
is consistent with the experimental upper limit found from $^{18}$F
parity-violating measurements.

\newpage
In this Letter, we use the method of QCD sum rules with the electroweak and
QCD Lagrangians to predict the weak parity-violating (PV)
pion-nucleon coupling constant, $f_{\pi NN}$.  The theoretical prediction
of $f_{\pi NN}$ is an important and challenging problem.
To-date, the most accurate PV experiments have only shown$^{1,2)}$ that the
upper limit for the magnitude of this coupling constant is 3-5 times smaller
than the ``best value'' predicted by DDH$^{3)}$ on the basis of a quark
model and somewhat smaller than that in a similar calculation carried out more
recently.$^{4)}$
Since that time others have tried to estimate $f_{\pi NN}$ by means
of  chiral soliton models$^{5,6)}$ and QCD sum rules.$^{7)}$
This coupling is of particular
interest because of its sensitivity to the neutral currents contribution
of weak nonleptonic processes at low energies.$^{2)}$

QCD sum rules have been shown to be able to reproduce known properties of the
nucleon, e.g., $\mu_p$, $\mu_n$, $g_A$, and of other hadrons.$^{8)}$
However, they have rarely (if ever) been used to predict unknown properties.
Keeping terms in the operator product expansion (OPE) up to dimension 5, we
show that there are two main terms in the sum rule for $f_{\pi NN}$:
the unit operator and a dimension D=3
susceptibility. By using an analogous sum rule for the strong coupling
constant, $g_{\pi NN}$, to evaluate this susceptibility, we are able to
determine the weak coupling $f_{\pi NN}$. An important aspect of the
present work is that we demonstrate that there is a cancellation between
perturbative and nonperturbative QCD modifications of the weak process.

We employ a two point function for the nucleon in an external pionic field.
Our current is the usual one$^{9)}$
\begin{eqnarray}
\eta_p (x) & = &\epsilon^{abc}[u^{aT}(x)C\gamma_\mu u^b(x)]\gamma^5
\gamma^\mu d^c(x),\nonumber \\
\bar\eta_p (y) & = &\epsilon^{abc}[\bar u^b(y)\gamma_\nu C\bar u^{aT}(y)]
\bar d^c(y) \gamma^\nu\gamma^5\;,
\end{eqnarray}
where $\epsilon^{abc}$ is the antisymmetric tensor, $C$ is the charge
 conjugation
operator, and $a,b,c$ are color indices.  The neutron currents are similar,
with the interchange of $d\leftrightarrow u$.

Since the most general weak PV $\pi$-$N$ coupling is$^{2,3,10)}$
\begin{equation}
H_{PV}(\pi NN) = {f_{\pi NN}\over \sqrt2} \bar\psi(\tau\times\phi_\pi)_3
\psi \;,
\end{equation}
only charged pions can be emitted or absorbed.  For definiteness, we
consider the absorption of a $\pi^+$ so that an initial neutron is converted
to a proton, and the correlator we consider is
\begin{equation}
\Pi = i\int d^4x e^{ix\cdot p}<0|T[\eta_p(x)\bar\eta_n(0)]|0>_{\pi^+}\;.
\end{equation}
The general form of $\Pi$ for the parity-violating pion-nucleon coupling,
as dictated by relativistic invariance, is
\begin{equation}
\Pi^{PV} = \Pi_e\;1+\Pi_o\hat p\;,
\end{equation}
with $\hat p \equiv \gamma_\mu p^\mu$.

The phenomenological evaluation of the correlator is carried out by deriving
a dispersion relation for $\Pi$ through the insertion of a complete set of
physical intermediate states of spin ${1\over 2}$ in the expression of Eq. 3.
Using the usual terminology, we refer to this as the right-hand side (RHS).
One finds for the parity-violating part of Eqs. 3,4:
\begin{equation}
\Pi_e^{PV}(p^2)^{RHS} = {\lambda_N^2 f_{\pi NN}(p^2+M^2)\over (p^2-M^2)^2} +
continuum.
\end{equation}
The double pole term corresponding to the insertion of the one-nucleon
intermediate state in Eq. (3) has contributions both from the weak pion-nucleon
vertex and the parity violation in the nucleon state itself. As will be shown
below, in our microscopic calculation using the two-point form, only the latter
process contributes.
The parameter $\lambda_N$ is related to the amplitude for finding three quarks
in a nucleon at one point and has been determined in a number of sum-rule
calculations.$^{8)}$  $M$ is the nucleon mass.  As is usual in the method, the
physical property of interest, $f_{\pi NN}$, is obtained by treating the
double-pole term explicitly, while the continuum and excited states are
included in the
numerical analysis via a parameterization, as discussed below.

The microscopic evaluation of $\Pi$, based on QCD and electroweak theory (the
so-called left-hand sides (LHS) of the QCD sum rules for $\Pi$), is obtained
by means of a Wilson coefficient expansion in inverse powers of $p^2$.  In
this work we keep diagrams up to dimension D = 5.
The lowest dimensional diagrams which we consider are shown in Fig. 1.
The higher dimensional diagrams which we include are obtained from those
shown in Fig. 1 by the substitution of Figs. 2b and c for the pion-quark
vertex, Fig. 2a. The propagators corresponding to the three diagrams of
Fig. 2 are:
\begin{eqnarray}
S^{ab}_{5a} & = &{i\vec\tau\cdot\vec\pi\over 4\pi^2x^2}\;g_{\pi q}
\gamma^5\delta^{ab} \nonumber \\
S^{ab}_{5b }& = & {i\over 12}\;\vec\tau\cdot\vec\pi\;g_{\pi q}\chi_\pi<\bar
qq>\delta^{ab}\gamma^5\;, \nonumber \\
S^{ab}_{5c} & = & -{i\over 3\cdot 2^6}\;m_{0}^{\pi}<\bar qq>g_{\pi q}
\vec\tau\cdot\vec\pi x^2 \gamma^5\;,
\end{eqnarray}
with $\chi_\pi g_{\pi q}<\bar qq>\equiv <\bar qi\gamma_5q>_\pi$ and
$<\bar q\;i\gamma_5 g_c \sigma\cdot Gq>_{\pi}\equiv m_{0}^{\pi}<\bar qq>
\vec\tau\cdot\vec\pi$.  Here $g_{\pi q}$ is the pion-quark coupling,
which is not explicitly used in the present calculation, and G represents
the gluon field.  The
susceptibility $\chi_\pi$ enters in the evaluation of both strong and weak
pion-nucleon coupling constants, while $m^\pi_0$ enters only for the weak one.
We will discuss the treatment of these parameters below.
We only consider the even sum rule, namely that for $\Pi_e$; that for $\Pi_o$
involves further unknown susceptibilities.
The evaluation of the diagrams is straightforward.

For the weak
Hamiltonian, we take $H_w = {G_F\over\sqrt2}\;(J^\mu J^\dagger_\mu
 + N^\mu N^\dagger_\mu)$ with
\begin{eqnarray}
J^\mu & = & \bar u\gamma^\mu(1-\gamma_5)d\cos\theta_C \nonumber \\
N^\mu & = &\bar u\gamma^\mu(A_u+B_u\gamma^5)u+
\bar d\;\gamma^\mu(A_d+B_d\gamma^5)d\;,
\end{eqnarray}
where $\theta_C$ is the Cabibbo angle and $A_u,A_d,B_u,B_d,$ are given by
\begin{eqnarray}
A_u & = & {1\over 2}(1-{8\over 3}\;\sin^2\theta_W), \nonumber \\
A_d& = & -{1\over 2} (1-{4\over 3}\;\sin^2\theta_W), \nonumber \\
B_u & = & -B_d = - {1\over 2}\;,
\end{eqnarray}
with $\theta_W$ the Weinberg angle.  This is the standard model Hamiltonian,
which we use for the main part of the calculation.  We then discuss the QCD
effects on our results.

Since momentum can be transferred in the weak point-like interaction, shown
by wavey lines representing Z$^0$ in the figures,
there is
an additional integral to be carried out in the evaluation of $\Pi$.  For
example, we obtain for Fig. 1a
\begin{equation}
\Pi_e^{1a} = {12 i\;\over \pi^{10}}\;G_F g_{\pi q}\int {(\hat x-\hat y)
\hat y y\cdot (x-y)\over (x-y)^8y^8x^2}(A_dB_u - A_u B_d)e^{ip\cdot x}
d^4xd^4y\;,
\end{equation}
There is no PV contribution from Figs. (1c) and (1d), and the sum of Figs.
(1b) and (1e) vanish.  The integrals in Eq.~(9) are evaluated by
standard Feynman techniques, with dimensional regularization.  The result
is
\begin{eqnarray}
\Pi_e^{1a}(p^2) & = & {-12\;G_Fg_{\pi q}\over \pi^8} \lim_{\epsilon\to 0}
\int
{e^{ip\cdot x}d^4x\over x^{10+2\epsilon}}\left(-{2\over \epsilon} +
2\gamma - {22\over
3}\right)(A_d B_u-A_u B_d) \nonumber \\
& = & {G_F\over 3\cdot 2^{8}\pi^6}\;g_{\pi q}\left({22\over 3}
- 2\gamma +
{2\over
\epsilon}\right)p^6\ln (-p^2)(A_dB_u-A_uB_d)\;,
\end{eqnarray}
where the infinite term $\propto \epsilon^{-1}$ is assumed to be regularized
away, and $\gamma$ is the Euler constant.  Taking the Borel transform one
obtains
\begin{equation}
\Pi_e^{1a} = -\;g_{\pi q}\;{ G_F\sin^2\theta_W\over 3\cdot 2^{6}\pi^6}
(11/3 - \gamma)\;M^8_B,
\end{equation}
where $M_B$ is the Borel mass.  The other diagrams can be evaluated in the
same manner.  The results from the processes of Figs. 1, and those obtained
from Figs. 1 with the substitution of Figs. 2b and c for Fig. 2a are
\begin{eqnarray}
\Pi_e^{PV}(M_B^2) & = & {G_F\sin^2\theta_W({11\over 3} - \gamma)
\over 12\pi^2}\;M^4_B\bigl[M^4_BL^{-4/9}E_3 \nonumber \\
& + &  {4\over 3}\chi_\pi\;aL^{-4/9} M^2_B E_2 - m_0^\pi\;a\;E_1
 L^{-4/9}\bigr]
\nonumber \\& = & f_{\pi NN}\bar\lambda^2_N\;e^{-M^2/M^2_B} (2{M^2\over
M_B^2} -1)
\end{eqnarray}
where $a=-(2\pi)^2<\bar qq>$ and $\bar\lambda^2_N$ = $(2 \pi)^4
\lambda^2_N/g_{\pi q}$. The lowest dimension pion-quark vertex renormalization
diagrams for $f_{\pi NN}$ are shown in Fig. 3a. In our approximation of a
contact weak interaction, the contribution of Figs. 3a-c vanish under a Borel
transformation. The mechanism of Figs. 3d and e do not appear in the external
field method. We do not include gluon condensate diagrams for $f_{\pi NN}$;
they are of the same order or smaller than uncertainties of our calculation.
The factors containing L, $L = 0.621\ln(10M_B)$, give the evolution in $Q^2$
arising from the anamolous dimensions, and the $E_i(M^2_B)$ functions take into
account excited states to ensure the proper large-$M_B^2$ behavior. The last
line in Eq. (12) is the Borel transform of the double-pole term from the
phenomenological (right-hand) side, Eq. 5. The direct proportionality to
$\sin^2\theta_W$ should be noted.

Finally, by explicit calculation or Fierz reordering,
we can show that the contribution
for $W^\pm$ exchanges vanish.  Thus, as required by
symmetries$^{3,10)}$, we find no charged current contribution to the weak
PV pion-nucleon vertex; such a contribution requires strangeness-changing
currents and would thus be reduced by $\sin^2\theta_C\approx 0.05$. Since we
neglect strangeness in the nucleon and strangeness-changing currents, we
obtain no contribution.

As we shall demonstrate below, the first two terms in the theoretical form
for $\Pi_e$ given in Eq. 12 are of opposite sign and tend to cancel. This
is a crucial point.  For this reason it is essential to either determine the
value of the susceptibility $\chi_\pi$ from $g_{\pi NN}$ or to eliminate it
from our equations.
We do both as an aid in determining the stability of our solutions.
First, we determine $\chi_\pi$ directly in terms of $g_{\pi NN}$ [as a
function of the Borel mass] by using the sum rule for the strong coupling,
which is analogous to Eq. 12, and attempt to use the result to determine
$f_{\pi NN}$.  Second, we eliminate
$\chi_\pi$ from the PV and strong coupling sum rules and find that we can
determine $f_{\pi NN}$ in terms of $g_{\pi NN}$.  Details are given below.

We use the correlator given by the two-point function of Eq. 3 for the strong
as well as the weak interaction.  The general form differs from Eq. 4 by the
presence of a $\gamma^5$ in each term.  The phenomenological (RHS) for the
strong pion-nucleon coupling is now given by
\begin{eqnarray}
\Pi_e^{s}(p^2)^{RHS} & = & {\lambda_N^2 g_{\pi NN} M^2\over (p^2-M^2)}\
\gamma^5 + continuum.
\end{eqnarray}
Unlike the weak PV pion-nucleon coupling, the evaluation of
the strong one leads to a problem in that there is no double pole on the
right-hand (dispersion relation) side.  However, as shown by Reinders et.
al.$^{11)}$, the
value of the coupling constant $g_{\pi NN}$ found in this way is
virtually
the same as that found by means of a 3-point function, which circumvents the
lack of a double pole problem.

Keeping terms up to D=6, shown in Fig. 4, for the theoretical side (LHS),
and taking the Borel transform we obtain the sum rule for
the strong pion-nucleon coupling:
\begin{eqnarray}
g_{\pi NN}\bar\lambda^2_N\;e^{-M^2/M^2_B} & = &M_B^6 L^{-4/9}E_2 -
M_B^4\chi_{\pi} a L^{2/9}E_1 +{4\over 3} a^2 L^{4/9} \nonumber\\
&&+ {<g_c^2 G^2>E_0M_B^2\over 8} - <g_c^2 G^2> E_0 M_B^2 ({13\over 8} -
 ln\,M^2),
\end{eqnarray}
where $<g_c^2 G^2>$ is the gluonic condensate.

Before we discuss our detailed evaluation of the sum rules to obtain our
estimate of $f_{\pi NN}$, let us discuss the structure of Eqs.~(12) and (14).
First, as we discuss below, if
we use PCAC to evaluate $\chi_{\pi}$ we find that $\chi_{\pi} a$=-44 GeV$^2$.
With this value, the $\chi_{\pi}$ term dominates both Eqs. (12) and (14), with
the result that $g_{\pi NN} \simeq$ 155 [in contrast to the experimental
value of 13.5].  With this value of $\chi_{\pi}$ we find that
$f_{\pi NN}~ \geq~ 10^{-6}$, at least an order of magnitude
larger than experiment.

Secondly, since $\chi_{\pi}$ is the only unknown in Eq. (14),
we can estimate the vacuum susceptibility using the
experimental value of $g_{\pi NN}$ = 13.5: this gives
$\chi_\pi\, a \simeq$-0.94 GeV$^2$, two orders of magnitude smaller that the
PCAC value [see discussion below].
With this value one finds that the
first two terms in Eq. (12), the leading terms for $f_{\pi NN}$, almost
cancel.  Note that the second term involving $\chi_{\pi}$ enters with
the opposite sign in the two equations for $f_{\pi NN}$ and $g_{\pi NN}$,
respectively.
This is the source of the very small parity-violating pion-nucleon
coupling in comparison with quark model: there is a cancellation between
the dimension zero model-like term using perturbative quark propagators
and the vacuum pion susceptibility term.

However, we find that the sum rule obtained for $f_{\pi NN}$ [Eq. (12)], using
the value of $\chi_\pi$(M$_B^2$) extracted from Eq. (14), is not stable in
M$_B$. Therefore we cannot obtain a reliable estimate of $f_{\pi NN}$ by this
method.

We find that we can obtain a satisfactory sum rule to determine
$f_{\pi NN}$ by eliminating $\chi_\pi$ from both
Eq. 12 and Eq. 14 by taking derivatives with respect to $M_B^2$.
With this procedure, and taking the ratio of the weak to the strong
sum rule we obtain the new sum rule for the weak in terms of the strong
coupling constant:
\begin{eqnarray}
{f_{\pi NN}\over g_{\pi NN}} & = & c_w M_N^2 {(M_N^2 - 4 M_B^2)
 (E_3 M_B^4- a m_0^\pi E_1)
\over (2 M_N^4 + 3 M_B^4 - 9 M_N^2 M_B^2) (12 E_2 M_B^4 +3 <G^2>E_0)},
\end{eqnarray}
where $c_w = G_F sin^2\theta_W({11\over 3} - \gamma)/(12 \pi^2)=6.7 \times
10^{-8}\, GeV^{-2}$.
The sum rule is quite stable with a plateau in M$_B^2$ in the region expected,
as shown in Fig. 5. Because of the strong cancellation between the first two
terms in Eq. (12) [dimension 0 and dimension 2 terms], the dimension four
term with the unknown parameter m$_0^\pi$ is important for the final
numerical value of $f_{\pi NN}$.  We have taken m$_0^\pi$ = 0 in Fig 5.
Guided
by the value of the parameter m$_0$ needed in the nucleon sum rule$^{8)}$, we
evaluate the sum rule given in Eq. 15 with m$_0^\pi$ taken over the range
0.0 to +0.8 GeV.
{}From this procedure we find:
\begin{eqnarray}
f_{\pi NN} & \approx & (2.3~ to~ 5.8) \times 10^{-8} for
 \nonumber\\
  m_0^\pi & = & (0~ to~ 0.8) GeV.
\end{eqnarray}
For negative values of m$_0^\pi$ the value of $f_{\pi NN}$ becomes smaller
and even negative, but we did not find stable solutions for sizable negative
values of this unknown parameter.
To be consistent with the neglect of gluon condensate terms we quote as
our central value of $f_{\pi NN}$ that with  $m_0^\pi = 0 $, shown in Fig. 5,
as
\begin{equation}
f_{\pi NN} \approx 2.3 \times 10^{-8}.
\end{equation}
This coupling constant is an order of magnitude smaller than the ``best values"
of Refs. 3 and 4.  As emphasized earlier, this result follows from
the cancellation of the two leading terms in Eq. 12.
The first LHS term in that equation,
a unit dimension term which would correspond to a quark
model type calculation gives, a value for $f_{\pi NN} \approx 2\times 10^{-7}$,
similar to the quark model value.  The second term, involving the
nonperturbative QCD vacuum susceptibility, $\chi_\pi$, strongly cancels
the first term.
Because of this cancellation, we cannot expect Eq. 17 to be very accurate,
but we find a clear explanation for the small value of $f_{\pi NN}$,
consistent with experiment.$^{1,2)}$

The results given in Eqs. 16 and 17 have been obtained using the Hamiltonian
of the standard model (see Eqs. 7 and 8).  Let us now consider the
strong interaction modifications.  These have been estimated in Refs. 3 and
4 using the renormalization group method.
In the notation of Ref. 3, the operators
involved in our calculation are $O_4$ and $O_5$.  Using the tables in
Refs. 3 and 4 we find that the our parameter for the parity
violation, $A_dB_u-A_uB_d$, would be changed by less than a factor of two
in magnitude.
Since the same parameter appears in all terms, this gives the overall
uncertainty arising from strong interaction modifications.  Therefore,
the main conclusion of our work is not changed.

There are two relevant features that we would like to point out.  The first
one is that the use of pseudovector coupling also circumvents the problem
of a lack of double pole for the strong interaction constant.
For the Lagrangian
\begin{equation}
{\cal L}_{\pi NN} = {g'_{\pi NN}\over
 m_\pi}\;\bar\psi_N\;i\gamma^\mu\gamma^5\vec
\tau\cdot\psi_N\nabla_\mu\vec\phi_\pi
\end{equation}
we can treat $\nabla_\mu\phi_\pi$ as a constant external axial vector field.
The QCD sum rule is then identical to our calculation of $g_A$.$^{8)}$  At the
quark level, we have
\begin{equation}
{\cal }L_{\pi qq} = {1\over 2f_\pi}\;\bar\psi_q\;i\gamma^\mu
\gamma^5\vec\tau\psi_q\nabla_\mu\vec\phi_\pi\;,
\end{equation}
where $f_\pi$ is the pion decay constant.  From our previous result for
$g_A$,$^{1)}$ we then obtain
\begin{eqnarray}
{g'_{\pi NN}\over m_\pi} & = & {g_A\over 2f_\pi}\;,\nonumber \\
g_{\pi NN} & = & g'_{\pi NN}\;{2M\over m_\pi} = {g_AM\over f_\pi}
\end{eqnarray}
which is just the Goldberger-Treiman relation.

As a second feature we wish to attempt an independent estimate of
$\chi_\pi$.  For this purpose we first use PCAC to obtain
\begin{equation}
<0|\bar u\;i\gamma_5u - \bar d\;i\gamma_5d|\pi^0> = {-f_\pi m^2_\pi\over
\sqrt2 m_q}\;e^{-iq\cdot x}\;.
\end{equation}
We then use the work of Belyaev and Kogan$^{12)}$, which assumes saturation
of a sum by one pion states:
\begin{eqnarray}
<0|\bar q\;i\gamma_5\tau_3 q|0>_\pi & = & {-i\over\sqrt2}\;\phi_\pi\;
\int d^4xe^{iQ\cdot x}<0|\bar u\;i\gamma_5 u -\bar d\;i\gamma_5
d|\pi><\pi|\bar q\;i\gamma_5\tau_3 q|0>_{Q\to 0} \nonumber \\
& = & {i\over\sqrt2}\phi_\pi\;{f^2_\pi m^2_\pi\over 2m^2_q}\equiv \chi_\pi
\phi_\pi <\bar qq>
\end{eqnarray}

As described above, the value of $\chi_\pi$ obtained in this manner is more
than an order of magnitude larger than that found by using the value of
$g_{\pi NN}$ from experiment.  Once more we point out that
if we use it in Eq. 14 we find an
order of magnitude discrepancy with the strong coupling constant,
$g_{\pi NN}$.
Furthermore, it is clear that this value of $\chi_\pi$
is inconsistent with Eq.~(12), since by eliminating it with derivatives
with respect to the Borel mass we obtain results an order of magnitude
different than with its use. We conclude that Eq.~(21)
cannot be correct.  We are not certain where the method of Belyaev and
Kogan errs, but we believe that it is suspect.

In conclusion, we find that
the weak PV pion-nucleon coupling due to neutral currents is as small
as that due to charged currents, $\sim 3 \times 10^{-8}$.  This
result agrees with the conclusion of the chiral soliton model of Kaiser and
Meissner$^{5)}$, but not that of Kaplan and Savage$^{6)}$. Our result also
disagrees with quark model calculations$^{3,4)}$ and with a previous QCD sum
rule calculation.$^{7)}$  If the coupling
is as small as we estimate, it cannot be separated from the charged current
contribution and thus cannot be found experimentally.
Although we have omitted gluon condensate corrections to the PV correlator, our
result is sufficiently small that these corrections will not alter our
conclusion. Finally, we point out that in the two-point QCD sum rule method
used here, the small value of $f_{\pi NN}$ which we obtained is the
result of a cancellation between a process which can be treated in quark models
and a vacuum process identified in the method of QCD sum rules.

\vskip 1 true cm
\centerline{\bf Acknowledgements}
\bigskip
The work of W-Y.P.H. was supported in part by the National Science Council of
R.O.C. (NSC84-2112-M002-021Y). The work of E. M. H. was supported in part by
the U.S. Department of Energy under grant DE-FG06-88ER40427,
while that of L.S.K. was supported in part by
the National Science Foundation grant PHY-9319641.
This work was also supported by the N.S.C. of
R.O.C. and the National Science Foundation of U.S.A. as a cooperative
research project. We would like to thank M. Savage for a helpful conversation.

\pagebreak

\begin{enumerate}

\item C.A. Barnes {\it et al.} Phys. Rev. Lett. \ut{40} (1978) 840;
H.C. Evans {\it et al.}, Phys. Rev. Lett. \ut{55} (1985) 791; Phys. Rev.
C\ut{35} (1987) 1119; M. Bini {\it et al.}, Phys. Rev. Lett. \ut{55}
(1985) 795; Phys. Rev. C\ut{38} (1988) 1195.

\item See e.g. E.G. Adelberger and W.C. Haxton, Ann. Rev. Nucl. Part.
Sci. \ut{35} (1985) 501; and J. Lang {\it et al.}, Phys. Rev. \ut{34} (1986)
1545.

\item B. Desplanques, J.F. Donoghue, and B.R. Holstein, Ann. Phys. (NY)
\ut{124} (1980) 449.

\item V.M. Dubovik and S.V. Zenkin, Ann. Phys. (NY) \ut{172} (1986) 100.

\item N. Kaiser and U.G. Meissner, Nucl. Phys. A\ut{499} (1989) 699;
A\ut{510} (1990) 1648 and U. Meissner, Mod. Phys. Lett. \ut{A5} (1990) 1703.

\item D.M. Kaplan and M.J. Savage, Nucl. Phys. A\ut{556} (1993) 653.

\item V.M. Khatsimovskii, Yad. Fiz. \ut{42} (1985) 1236 [transl. Sov. J.
Nucl. Phys. \ut{42} (1985) 781].

\item For references to early work see L.J. Reinders, H. Rubinstein and
S. Yazaki, Nucl. Phys. B\ut{213} (1983) 109.
For references to more recent work see, e.g., E.M. Henley and J. Pasupathy,
Nucl. Phys. A\ut{556} (1993) 467; E.M. Henley, W-Y.P. Hwang and L.S.
Kisslinger, Phys. Rev. D\ut{46} (1992) 431;
T. Hatsuda {\it et al.}, Phys. Rev. C\ut{49} (1993) 452.

\item B.L. Ioffe, Nucl. Phys. B\ut{188} (1981) 317; B\ut{191} (1981) 591(E); Z.
Phys. C\ut{18} (1983) 67.

\item E.M. Henley, Ann. Rev. Nucl. Sci. \ut{19} (1969) 367; Chinese J.
Phys. \ut{30} (1992) 1.

\item L.J. Reinders, H.R. Rubinstein, and S. Yazaki, Nucl. Phys. B\ut{213}
(1983) 109; L.J. Reinders, Acta Phys. Polon. B\ut{15} (1984) 329.

\item V.M. Belyaev and Ya. I. Kogan, Phys. Lett. \ut{136B} (1984) 273.

\end{enumerate}
\centerline{\bf Figure Captions}
\bigskip
\parindent=0truept
Fig. 1. Lowest dimension quark diagrams for the PV weak pion-nucleon vertex.
The dashed line represents a charged pion and the wavy line a $Z^0$.

\medskip
Fig. 2. Quark propagator modifications in an external pion field.

\medskip
Fig. 3. Pion-nucleon weak vertex correction diagrams.

\medskip
Fig. 4. Diagrams contributing to the calculation of $g_{\pi NN}$.

\medskip
Fig. 5. Solution for $f_{\pi NN}$ from Eq. 15 as a function of M$_B$.
\end{document}